\documentstyle[11pt,epsfig]{article}
\begin{document}

\title{
A perturbative approach for the dynamics of the quantum Zeno
subspaces}

\author{Yi-Xin Chen \thanks{Email:yxchen@zimp.zju.edu.cn}
\hspace{.5mm} and Zhuowen Fang  \\[.3cm]
{\small Zhejiang Institute of Modern Physics, Zhejiang University,}\\
             {\small Hangzhou 310027, P.R.China,}}

\date{\today}
\maketitle

\begin{abstract}

\indent

In this paper we investigate the dynamics of the quantum Zeno
subspaces which are the eigenspaces of the interaction
Hamiltonian, belonging to different eigenvalues. Using the
perturbation theory and the adiabatic approximation, we get a
general expression of the jump probability between different Zeno
subspaces. We applied this result in some examples. In these
examples, as the coupling constant of the interactions increases,
the measurement keeps the system remaining in its initial subspace
and the quantum Zeno effect takes place.
\\
\\
PACS numbers: 03.65.Xp, 03.65.Ta, 03.67.Lx

\vspace{.5cm}

\end{abstract}

\setcounter{equation}{0}

\section{Introduction}

\indent

The quantum Zeno effect \cite{Khalfin}\cite{Fonda} has attracted
great attentions. This phenomenon is caused by the influence of
the measurement on the evolution of a quantum system. Frequent
measurements can inhibit the decay of any unstable system
\cite{Misra}, and the short time behavior of the survival
probability is not exponential but quadratic. The deviation from
the exponential decay has been confirmed in a tunnelling
experiment by wilkinson et al \cite{Wilkinson}. Moreover, it was
also predicted that frequent measurements(but not too frequent)
could accelerate the decay process. This is so-called quantum
anti-Zeno effect. The quantum Zeno effect and anti-Zeno effect had
been discussed in ref.\cite{Kofman}\cite{Ruseckas}. Both effects
were first observed recently in an atomic tunnelling system
\cite{Fisher}.

Misra and Sudarshan's theorem \cite{Misra} proved that a system
was forced to evolve inside a subspace, related to a projection
operator, by frequently observations, but not remaining in its
initial state which belonged to the subspace. This idea was
developed by Facchi et al to frame the Zeno dynamics of a whole
system including a detector apparatus \cite{Facchi}. The system
can just evolve in a set of orthogonal subspaces of the total
Hilbert space which belong to different eigenvalues of the
interaction Hamiltonian in the infinitely strong coupling limit.
These subspaces, which the measurement process is able to
distinguish, are called quantum Zeno subspaces.

Quantum zeno dynamics is not absolutely developed yet. Up to now,
The dynamics of quantum Zeno subspaces in ref.\cite{Facchi} has
discussed the ``infinitely strong measurement" limit. But the
finitely strong measurement is untouched. In this paper, we
combine perturbation theory and adiabatic approximation to
describe such process. We obtain an expression for the jump
probability between two different Zeno subspaces of the
interaction Hamiltonian. Therefore we have a general method to
deal with the dynamics of quantum Zeno dynamics.

The organization of this paper is as follows. In Section 2 we
briefly review the quantum Zeno subspace theorem. In section 3, we
apply the perturbative method to get a general express of the jump
probability between different Zeno subspaces of the interaction
Hamiltonian. This method is unlike in ref.\cite{Ruseckas}. we
regard the free Hamiltonian of the measured system and the
detector apparatus as a perturbation of the Hamiltonian describing
the interaction of them. We use this expression to analyze two
time-independent measurements in Section 4. We also use it to show
the QZE by performing the measurement on Heisenberg spin chain in
Section 5. The Section 6 concludes the summary of our results and
the discussions for them.

\section{Quantum Zeno subspaces}

\indent

We briefly introduce the quantum Zeno subspace theorem. This
theorem is developed by Facchi et al \cite{Facchi}.

Consider a quantum system described by the Hamiltonian $H$:
\begin{equation}
H=H_{0}+H_{meas}(K)
\end{equation}
Here $H_{0}$ is the free Hamiltonian of the measured system and
the detector apparatus, $H_{meas}(K)$ denotes the interaction of
them and $K$ is a set of coupling parameters. If $K$ is simply a
coupling constant, we can simplify the above Hamiltonian into the
form
\begin{equation}\label{H}
H=H_{0}+KH_{meas}
\end{equation}
The evolution of the system is described by the unitary operator
$U(t)$, which is completely determined by the total Hamiltonian
$H$. In the $K\rightarrow\infty$ limit, benefit from the adiabatic
theorem \cite{Facchi}\cite{Messiah}\cite{Born}, the evolution
operator
\begin{equation}
u(t)=\lim_{K\rightarrow\infty}U(t)
\end{equation}
has the property:
\begin{equation}\label{u}
[u(t),P_{n}]=0,
 \end{equation}
where
\begin{eqnarray}
H_{meas}P_{n}=\varepsilon_{n}P_{n}, \quad P_{n}P_{m}=\delta_{nm}.
 \end{eqnarray}
$P_{n}$ is a orthogonal operator which projects the total Hilbert
space onto ${\cal H}_{P_{n}}$, the eigenspace of $H_{meas}$
belonging to the eigenvalue $\varepsilon_{n}$. These subspaces are
called quantum Zeno subspaces. If the eigenvalue is degenerate,
the corresponding quantum Zeno subspace is the plus of the
degenerate eigenspaces. Therefore they are in general
multidimensional. We can see that the operator $u(t)$ is diagonal
with respect to $H_{meas}$. Moreover, if $H$ is time-independent,
the evolution operator $u(t)$ can be explicitly given by
\begin{equation}
u(t)=\exp\{-i\sum_{n}(P_{n}H_{0}P_{n}+K\varepsilon_{n}P_{n})t\}.
\end{equation}

Let the system in the initial density matrix $\rho_{0}$. In the
$K\rightarrow\infty$ limit, the density matrix of the system is
\begin{equation}
\rho(t)=u(t)\rho_{0}u^{\dagger}(t),
\end{equation}
and the probability to find the system in ${\cal H}_{P_{n}}$ is
\begin{eqnarray}
p_{n}(t)=Tr[\rho(t)P_{n}]=Tr[u(t)\rho_{0}u^{\dagger}(t)P_{n}]
        =Tr[u(t)\rho_{0}P_{n}u^{\dagger}(t)]
        =Tr[\rho_{0}P_{n}]
        =p_{n}(0).
\end{eqnarray}
From this result, it is clear to see that the probability in each
quantum Zeno subspace does not change during the measurement
process. If the initial density matrix belongs to a quantum Zeno
subspace
\begin{equation}
\rho_{0}=P_{n}\rho_{0}P_{n},
\end{equation}
 the system will remain there forever and the
QZE takes place.

In $K\rightarrow\infty$ limit, the interaction Hamiltonian plays
the leading role and determines the evolution of the system. Each
quantum Zeno subspace evolves individually, so the probability of
each subspace does not leak out to another, although the system
does not remain in its initial state.

\section{An approximate method }

\indent

In Sec.2, Quantum Zeno subspaces have been investigated in
$K\rightarrow\infty$ limit. On the other hand, we pay great
attention to the finitely strong time-dependent measurement. We
want to know the jump probability between different Zeno subspaces
and more details about the quantum Zeno effect. We find that if
the free Hamiltonian $H_{0}(t)$ compared with the interaction
Hamiltonian $KH_{meas}(t)$ is a perturbation and $KH_{meas}(t)$
satisfies the adiabatic approximation condition, the jump
probability is mainly from a contribution of the second-order
approximation of the density matrix.

\subsection{Perturbation theory}

\indent

We still consider the system whose time-dependent Hamiltonian has
the form (\ref{H}). The Hamiltonian $H_{0}(t)$ and $KH_{meas}(t)$
are Hermitian operators respectively. $H_{0}(t)$ is a perturbation
of $KH_{meas}(t)$. The time evolution operator $U(t,0)$ is
determined by the Schr\"{o}dinger equation
\begin{equation}
i\frac{d}{dt}U(t,0)=(H_{0}(t)+H_{meas}(t))U(t,0) ,\quad
U(0,0)=1.
\end{equation}
Here Plank's constant $\hbar$ equals 1. Let us review in summary
the solution of the general expression of $U(t,0)$ \cite{Messiah}.
We assume $U^{(0)}(t,0)$ is the unitary time evolution operator
corresponding to $KH_{meas}(t)$:
\begin{equation}
i\frac{d}{dt}U^{(0)}(t,0)=KH_{meas}(t)U^{(0)}(t,0),\quad
U^{(0)}(0,0)=1.
\end{equation}
We change the Schr\"{o}dinger representation into the intermediate
``representation" by the unitary transformation
$U^{(0)\dagger}(t,0)$:
\begin{equation}\label{U}
U_{I}(t,0)=U^{(0)\dagger}(t,0)U(t,0).
\end{equation}
The Schr\"{o}dinger equation in this ``representation" reads
\begin{equation}\label{new equation}
i\frac{d}{dt}U_{I}(t,0)=H_{I0}(t)U_{I}(t,0),
\end{equation}
where
\begin{equation}
H_{I0}(t)=U^{(0)\dagger}(t,0)H_{0}U^{(0)}(t,0).
\end{equation}
The formal solution of Eq.(\ref{new equation}) is
\begin{equation}
U_{I}(t,0)=T\exp(-i\int_{0}^{t}H_{I0}(t^{'})dt^{'}).
\end{equation}
Here T denotes the time-ordering. According to Eq.(\ref{U}), we get
the expansion for $U(t,0)$:
\begin{equation}
U(t,0)=U^{(0)}(t,0)+\sum_{n=1}^{\infty}U^{(n)}(t,0),
\end{equation}
where
\begin{eqnarray}\label{Un}
U^{(n)}(t,0)&=&(-i)^{n}\int_{0}^{t}dt_{1}\int_{0}^{t_{1}}dt_{2}\cdot\cdot\cdot\int_{0}^{t_{n-1}}dt_{n}U^{(0)}(t,t_{n})H_{0}(t_{n}) \nonumber \\
            && \times U^{(0)}(t_{n},t_{n-1})H_{0}(t_{n-1})\cdot\cdot\cdot
            U^{(0)}(t_{2},t_{1})H_{0}(t_{1})U^{(0)}(t_{1},t_{0}).
\end{eqnarray}
The expansion is power series in $H_{0}(t)$. If the measurement is
strong and $U^{(0)}(t,0)$ is very close to $U(t,0)$, the series
converge rapidly. In the first-order approximation, we have
\begin{eqnarray}\label{first-order}
U(t,0)=U^{(0)}(t,0)+(-i)\int_{0}^{t}dt_{1}U^{(0)}(t,t_{1})H_{0}(t_{1})U^{(0)}(t_{1},0).
\end{eqnarray}

Since $H_{meas}(t)$ is time-dependent, its eigenspaces can shift
during the measurement process, as well as the eigenvalues
$\varepsilon_{n}(t)$. We have
 \begin{eqnarray}
H_{meas}(t)P_{n}(t)&=&\varepsilon_{n}(t)P_{n}(t),\\
P_{n}(t)P_{m}(t)&=&\delta_{nm}.
\end{eqnarray}
The Hilbert space corresponding to the projection $P_{n}(t)$ is in
general multidimensional. We suppose at the initial time the
quantum system is in $\rho_{0}$. $\rho_{0}$ belongs to a quantum
Zeno subspace. It means
\begin{equation}
\rho_{0}=P_{n}(0)\rho_{0}P_{n}(0).
\end{equation}
Under the continuous measurement, the density matrix at time $t$
becomes
\begin{equation}
\rho(t)=U(t,0)\rho_{0}U^{\dagger}(t,0).
\end{equation}
Using Eq.(\ref{first-order}), the density matrix can be obtained
up to second-order approximation:
\begin{eqnarray}
\rho^{(0)}(t)&=&U^{(0)}(t,0)\rho_{0}U^{(0)\dagger}(t,0) \label{0} , \\
\rho^{(1)}(t)&=&U^{(0)}(t,0)\rho_{0}U^{(1)\dagger}(t,0)+U^{(1)}(t,0)\rho_{0}U^{(0)\dagger}(t,0),
\label{1} \\
\rho^{(2)}(t)&=&U^{(0)}(t,0)\rho_{0}U^{(2)\dagger}(t,o)+U^{(2)}(t,0)\rho_{0}U^{(0)\dagger}(t,0)
           +U^{(1)}(t,0)\rho_{0}U^{(1)\dagger}(t,0). \label{2}
\end{eqnarray}

\subsection{Adiabatic approximation}

\indent

The property Eq.(\ref{u}) \cite{Facchi} of the Quantum Zeno
subspaces is derived from the adiabatic theorem
\cite{Messiah}\cite{Born}. Similarly we apply the adiabatic
approximation to solve the time-dependent measurement problem.

Throughout the measurement process, we suppose the eigenvalues and
the eigenspaces of the interaction Hamiltonian $KH_{meas}(t)$
satisfy \cite{Messiah}:

(i) the eigenvalues remain distinct:
 \begin{equation}
\varepsilon_{n}(t)\neq\varepsilon_{m}(t),\quad m\neq n;
\end{equation}

(ii) the derivatives $dP_{n}(t)/dt$, $d^{2}P_{n}(t)/dt^{2}$ are
well-defined and piece-wise continuous.\\
We define a unitary operator A(t) having the property
\begin{equation}
P_{n}(t)=A(t)P_{n}(0)A^{\dagger}(t), \quad A(0)=1.
\end{equation}
The physical significance of the unitary transformation $A(t)$ is
that: it takes any set of basis vectors of $H_{meas}(0)$ over into
a set of basis vectors of $H_{meas}(t)$, each eigenvectors of
$H_{meas}(0)$ being carried over into one of the eigenvectors of
$H_{meas}(t)$ that derive from it by continuity. It is determined
by the following equation
\begin{equation}
i\frac{d}{dt}A(t)=M(t)A(t),
\end{equation}
where $M(t)$ is a Hermitian operator
\begin{equation}
M(t)=i\sum_{n}(dP_{n}(t)/dt)P_{n}(t).
\end{equation}

We assume that $KH_{meas}(t)$ satisfy the adiabatic approximation
condition
\begin{equation}
\mid\frac{\alpha_{m}^{max}}{\varepsilon_{m}^{min}}\mid\ll K^{2},
\end{equation}
where
\begin{equation}
\varepsilon_{m}^{min}=min|\varepsilon_{m}(t)-\varepsilon_{n}(t)|,
\quad m\neq n,
\end{equation}
and
\begin{eqnarray}
\alpha_{m}^{max}=max(\sum_{m\neq n}\mid\alpha_{mn}(t)\mid^{2}) \\
\alpha_{mn}(t)=-_{t}\langle m|\frac{dH_{meas}(t)}{dt}|
            n\rangle_{t}/\varepsilon_{mn}(t)
\end{eqnarray}
Here $|n\rangle_{t}$ is the initial eigenvector of $H_{meas}(t)$
belonging to the eigenvalue $\varepsilon_{n}(t)$ , $|m\rangle_{t}$
belonging to $\varepsilon_{m}(t)$. $\varepsilon_{mn}(t)$ is the
``Bohr frequency" of the transition n $\rightarrow$ m. Therefore
the zero-approximation $U^{(0)}(t,0)$ of the time evolution
operator determined by $KH_{meas}(t)$ has the asymptotic property
\begin{equation}\label{UP}
U^{(0)}(t,0)P_{n}(0)=P_{n}(t)U^{(0)}(t,0),
\end{equation}
and $U^{(0)}(t,0)$ can be expressed approximately in the form
\begin{equation}\label{U0}
U^{(0)}(t,0)\simeq A(t)\Phi(t),
\end{equation}
where
\begin{eqnarray}
\Phi(t)&=&\sum_{n}\exp(-i\varphi_{n}(t))P_{n}(0), \\
\varphi_{n}(t)&=&\int_{0}^{t}K\varepsilon(t^{'})dt^{'}.
\end{eqnarray}
Specially, if the interaction Hamiltonian is time-independent,
$A(t)$ equals 1 at any time, and $\Phi(t)$ is
\begin{equation}
\Phi(t)=\sum_{n}\exp(-iK\varepsilon_{n}t)P_{n}.
\end{equation}

\subsection{Jump probability}

\indent

We investigate the jump probability from Zeno subspace ${\cal H}_{
P_{n}(0)}$ to ${\cal H}_{P_{m}(t)}$ under the action of the
perturbation $H_{0}(t)$. The jump probability is
\begin{equation}\label{W}
W(P_{n}(0)\rightarrow P_{m}(t))=Tr\{P_{m}(t)\rho(t)\}.
\end{equation}
Since the initial density matrix belongs to the quantum Zeno
subspace ${\cal H}_{P_{n}}$, we have
\begin{eqnarray}\label{property}
P_{m}(0)\rho_{0}=\rho_{0}P_{m}(0)=0,\quad m\neq n.
\end{eqnarray}
Using Eq.(\ref{0}), (\ref{1}), (\ref{2}) and (\ref{UP}), we get
the expansion of the jump probability up to second-order. From
Eq.(\ref{property}), we find the zero-order term and first-order
term is 0 and the jump probability is mainly from the contribution
of the second-order term
\begin{eqnarray}
W^{(2)}(P_{n}(0)\rightarrow P_{m}(t))&=&Tr\{P_{m}(t)\rho^{(2)}(t)\}  \nonumber \\
                                     &=&Tr\{U^{(0)}(t,o)P_{m}(0)\rho_{0}U^{(2)\dagger}(t,o)\}+Tr\{U^{(2)}(t,0)\rho_{0}P_{m}(0)U^{(0)\dagger}(t,0)\}  \nonumber \\
                                      &&+Tr\{P_{m}(t)U^{(1)}(t,0)\rho_{0}U^{(1)\dagger}(t,0)\}\nonumber\\
                                     &=&Tr\{P_{m}(t)U^{(1)}(t,0)\rho_{0}U^{(1)\dagger}(t,0)\}.
\end{eqnarray}
Using Eq.(\ref{Un}), we get the probability defined by the
integral equation
\begin{equation}\label{P}
W(P_{n}(0)\rightarrow
P_{m}(t))=\int_{0}^{t}dt_{1}\int_{0}^{t}dt_{2}Tr\{P_{m}(t)U^{(0)}(t,t_{1})H_{0}(t_{1})U^{(0)}(t_{1},0)\rho_{0}
          U^{(0)\dagger}(t_{2},0)H_{0}(t_{2})U^{(0)\dagger}(t,t_{2})\}.
\end{equation}
From Eq.(\ref{U0}) and the composition law
\begin{equation}
U^{(0)}(t,0)=U^{(0)}(t,t^{'})U^{(0)}(t^{'},0),
\end{equation}
we can replace $U^{(0)}(t,t^{'})$ by the asymptotic form
\begin{equation}
U^{(0)}(t,t^{'})\simeq
A(t)\Phi(t)\Phi^{\dagger}(t^{'})A^{\dagger}(t^{'}).
\end{equation}
Eq.(\ref{P}) can be simplified to
\begin{eqnarray}\label{JP}
W(P_{n}(0)\rightarrow
P_{m}(t))&=&\int_{0}^{t}dt_{1}\int_{0}^{t}dt_{2}Tr\{A^{\dagger}(t_{1})H_{0}(t_{1})A(t_{1})\rho_{0}
          A^{\dagger}(t_{2})H_{0}(t_{2})A(t_{2})P_{m}(0)\} \nonumber \\
          &&\times
          \exp\{i\int_{t_{2}}^{t_{1}}K(\varepsilon_{m}(t^{'})-\varepsilon_{n}(t^{'}))dt^{'}\}.
\end{eqnarray}
There have two assumptions for the validity of Eq.(\ref{JP}): the
free Hamiltonian $H_{0}(t)$ can be regarded as a perturbation of
the interaction Hamiltonian $KH_{meas}(t)$ and $KH_{meas}(t)$
changes sufficiently slowly to satisfy the adiabatic approximation
condition. With the enhancement of the coupling constant $K$, the
phase factor vibrates rapidly and the integration tends to
decline. The decay of the system is inhibited by the measurement.
Eq.(\ref{JP}) is the main result of this paper. It can describe
the problem of the finitely strong measurement. We will discuss
the quantum Zeno effect in the following two Sections. However, in
the ``infinitely strong measurement" limit $K\rightarrow\infty$,
Eq.{\ref{JP}) tends to zero. Therefore the system remains in its
initial Zeno subspace ${\cal H}_{ P_{n}(0)}$ forever. This is the
result in ref.\cite{Facchi}.

\section{Time-independent measurement}

\indent

In the preceding section, we get the jump probability
Eq.(\ref{JP}) between different quantum Zeno subspaces. Now we use
it to look at time-independent measurement. Furthermore, we assume
the free Hamiltonian is time-independent. We consider the repeated
measurements separated by the free evolution of the system. The
duration of the free evolution is $\tau_{F}$ and the duration of
the measurement is ($\tau-\tau_{F}$):
\begin{equation}
H_{meas}(t)=\theta(\tau-t)\theta(t-\tau_{F})P_{n}.
\end{equation}
Here $\theta(t-\tau_{F})$ is Heaviside unit step function. There
are two Zeno subspaces ${\cal H}_{P_{n}}$ and ${\cal H}_{P_{m}}$
of $H_{meas}$ respectively belonging to the eigenvalues 1 and 0.
The initial density matrix $\rho_{0}$ of the system belongs to
Hilbert space ${\cal H}_{P_{n}}$. After a measurement, the jump
probability from ${\cal H}_{P_{n}}$ to ${\cal H}_{P_{m}}$ is
\begin{eqnarray}
W(P_{n}\rightarrow P_{m},\tau)&=&Tr\{P_{m}H_{0}\rho_{0}H_{0}\}[\tau_{F}^{2}+\frac{4\tau_{F}}{K}\sin\frac{1}{2}K(\tau-\tau_{F})\cos\frac{1}{2}K(\tau+\tau_{F}) \nonumber \\
      &&+\frac{4}{K^{2}}\sin^{2}\frac{1}{2}K(\tau-\tau_{F})].
\end{eqnarray}
In $\tau_{F}\rightarrow\tau$ limit(instantaneous measurement
\cite{von Neumann}), the survival probability exhibit a quadratic
behavior at short time:
\begin{equation}\label{WM}
W(P_{n},\tau)=1-\tau^{2}/\tau^{2}_{z},
\end{equation}
where
\begin{equation}
\tau^{-2}_{z}=Tr\{P_{m}H_{0}\rho_{0}H_{0}\}.
\end{equation}
$\tau_{z}$ is called Zeno time. We perform N measurements at time
intervals $\tau$ for a time $t$. With N increasing($\tau
\rightarrow 0$), the system will be freezed in its initial
subspace(QZE) \cite{Facchi}. This result is correct for the case
of the finite coupling constant $K$.

Let us now consider another time-independent continuous
measurement described by the following:
\begin{equation}
H_{meas}=\sum_{n}\varepsilon_{n}P_{n}.
\end{equation}
Similarly we have the initial density of the system belonging to
Hilbert space ${\cal H}_{P_{n}}$. The duration of the measurement
is $\tau$. The jump probability from Zeno subspace ${\cal
H}_{P_{n}}$ to ${\cal H}_{P_{m}}$ is
\begin{equation}
W(P_{n}\rightarrow
P_{m},\tau)=Tr\{P_{m}H_{0}\rho_{0}H_{0}\}\frac{4\sin^{2}\frac{1}{2}K(\varepsilon_{m}-\varepsilon_{n})\tau}{K^{2}(\varepsilon_{m}-\varepsilon_{n})^{2}}.
\end{equation}
And the survival probability is
\begin{eqnarray}
W(P_{n},\tau)&=&1-\sum_{m\neq n}W(P_{n}\rightarrow P_{m},\tau) \nonumber \\
          &=&1-\sum_{m\neq
           n}Tr\{P_{m}H_{0}\rho_{0}H_{0}\}\frac{4\sin^{2}\frac{1}{2}K(\varepsilon_{m}-\varepsilon_{n})\tau}{K^{2}(\varepsilon_{m}-\varepsilon_{n})^{2}}. \label{continuous}
\end{eqnarray}
When the system evolves under the continuous measurement for a
shot time $\tau$, we perform an ideal measurement(projection) to
confirm whether the system  survives inside ${\cal H}_{P_{n}}$.
Repeating the above procedure, we have the survival probability in
${\cal H}_{P_{n}}$ at time $t=N\tau$
\begin{eqnarray}
W(P_{n},t)&=&(W(P_{n},\tau))^{N} \nonumber \\
          &\simeq&1-t\sum_{m\neq n}Tr\{P_{m}H_{0}\rho_{0}H_{0}\}\frac{4\sin^{2}\frac{1}{2}K(\varepsilon_{m}-\varepsilon_{n})\tau}{K^{2}(\varepsilon_{m}-\varepsilon_{n})^{2}\tau} \label{continuous}\nonumber \\
          &\simeq&\exp(-Rt),
\end{eqnarray}
where the decay rate is
 \begin{equation}\label{R}
R=\sum_{m\neq
           n}Tr\{P_{m}H_{0}\rho_{0}H_{0}\}\frac{4\sin^{2}\frac{1}{2}K(\varepsilon_{m}-\varepsilon_{n})\tau}{K^{2}(\varepsilon_{m}-\varepsilon_{n})^{2}\tau}. \label{continuous}
\end{equation}
Introducing the functions
\begin{eqnarray}
G(\varepsilon)&=&\sum_{m\neq n}
Tr\{P_{m}H_{0}\rho_{0}H_{0}\}\delta(\varepsilon_{m}-\varepsilon), \\
F(\varepsilon)&=&\frac{4\sin^{2}\frac{1}{2}K(\varepsilon
-\varepsilon_{n})\tau}{2\pi
K^{2}(\varepsilon-\varepsilon_{n})^{2}\tau},
\end{eqnarray}
we can recast Eq.(\ref{R}) as
\begin{equation}\label{Re}
R=2\pi\int_{-\infty}^{+\infty}G(\varepsilon)F(\varepsilon)d\varepsilon.
\end{equation}
The above formulation is similar to the one obtained in
ref.\cite{Kofman} which has analyzed the conditions to obtain the
QZE and AZE. But the measurement process is completely different.
In that case, the free evolution of the system is interrupted by
instantaneous ideal measurements(projections) at time intervals
$\tau$. Furthermore, the measurements force the measured system
remaining in its initial state. In our case, the whole system
including the detector apparatus is not necessary to do that, but
remains in its initial Zeno subspace which is in general
multidimensional. The decay rate(\ref{Re}) is the overlap of the
factors $G(\varepsilon)$ and $F(\varepsilon)$. If the frequency
$\nu\sim 1/\tau$ satisfies
\begin{equation}
\nu\gg\Gamma_{R},|\varepsilon_{n}-\varepsilon_{M}|,
\end{equation}
the QZE can be obtained. Here $\Gamma_{R}$ is the width of
$G(\varepsilon)$ and $\varepsilon_{M}$ is the centre of gravity of
$G(\varepsilon)$. Moreover, the decay rate also reduces as the
coupling constant $K$ increases. In fact, the interaction
Hamiltonian which denotes the continuous measurement as well as
the free Hamiltonian govern the evolution of the system, the
bigger the more influence on it. We see that the evolution of the
system under the action of a continuous measurement process  is
similar to that obtained with pulsed measurements \cite{Facchi}.

\section{Measurement on Heisenberg spin chain}

\indent

Let us now investigate another example of a time-dependent
measurement on an XYZ Heisenberg spin(1/2) chain at zero
temperature. The spin-systems have been discussed in the subject
of Adiabatic Quantum Computation \cite{Murg}. The interesting
problem about the Adiabatic Quantum Computation is the
investigation of the ground state of spin-systems. In this paper,
we will investigate the quantum Zeno effect of the spin-systems
which initially is in the ground state. We have the spin chain
interacting with a magnetic field which is rotated sufficiently
slowly from Z-axis to X-axis without changing its magnitude $h$
during the time $T$ \cite{Korepin}. The total Hamiltonian of the
system is
\begin{equation}
H=H_{0}-\sum_{j=1}^{n}h((1-s)\sigma_{j}^{z}+s\sigma_{j}^{x}),\quad
s=\frac{t}{T}.
\end{equation}
Here $H_{0}$ represents the free Hamiltonian of the XYZ Heisenberg spin
chain
\begin{equation}
H_{0}=\sum_{j=1}^{n}( \lambda_{1}\sigma_{j}^{x}\sigma_{j+1}^{x}+\lambda_{2}\sigma_{j}^{y}\sigma_{j+1}^{y}+\lambda_{3}\sigma_{j}^{z}\sigma_{j+1}^{z} ),
\end{equation}
and the second-term denotes the interaction of the field and the
spin chain. We use the similar method in ref \cite{Korepin} by
Korepin to get the adiabatic approximation condition for the
interaction Hamiltonian $I(s)$:
 \begin{equation}\label{adiabatic}
  hT\gg\sqrt{n/2}.
  \end{equation}
This condition, unlike in ref \cite{Korepin}, is only for the
interaction Hamiltonian. The free Hamiltonian $H_{0}$ acts as a
perturbation.

We define
\begin{eqnarray}
I(s)&=&-\sum_{j=1}^{n}I_{j}(s) \nonumber \\
 &=&-\sum_{j=1}^{n}h((1-s)\sigma_{j}^{z}+s\sigma_{j}^{x}),
\end{eqnarray}
where
\begin{equation}\label{Ij}
I_{j}(s)=h((1-s)\sigma_{j}^{z}+s\sigma_{j}^{x}).
\end{equation}
Introducing the matrices
\begin{eqnarray}
A(s)&=&\pmatrix{\frac{s}{\sqrt{2K^{2}(s)-2K(s)(1-s)}} &
\frac{s}{\sqrt{2K^{2}(s)+2K(s)(1-s)}} \cr
\frac{K(s)-(1-s)}{\sqrt{2K^{2}(s)-2K(s)(1-s)}} &
\frac{-K(s)-(1-s)}{\sqrt{2K^{2}(s)+2K(s)(1-s)}} \cr },  \label{A}\\
A^{\dagger}(s)&=&\pmatrix{\frac{s}{\sqrt{2K^{2}(s)-2K(s)(1-s)}} &
\frac{K(s)-(1-s)}{\sqrt{2K^{2}(s)-2K(s)(1-s)}} \cr
\frac{s}{\sqrt{2K^{2}(s)+2K(s)(1-s)}} &
\frac{-K(s)-(1-s)}{\sqrt{2K^{2}(s)+2K(s)(1-s)}} \cr },\label{AD}
\end{eqnarray}
where
\begin{equation}
K(s)=\sqrt{s^{2}+(1-s)^{2}},
\end{equation}
we can rewrite the Eq.(\ref{Ij}) in the form:
 \begin{equation}\label{easy}
I_{j}(s)=hK(s)A_{j}(s)\sigma_{j}^{z}A_{j}^{\dagger}(s).
\end{equation}

We define the instantaneous eigenspaces of $I(s)$ is ${\cal
H}_{P_{Lm}}(s)$ corresponding to the projections $P(Lm,s)$. The
denotations of the number $L$ and $m$ are pointed out in the
following. At time $t=0$, the projection $P(Lm,0)$ is
\begin{equation}
 P(Lm, 0)=\bigotimes_{j=1}^{n}P_{j}(0).
 \end{equation}
Here $P_{j}(0)$ is the projection onto the eigenspace ${\cal
H}_{P_{j}}(0)$ of $I_{j}(0)$. From Eq.(\ref{easy}), we can easily
write the projection at time $s$ in the form:
\begin{equation}
P(L m,s)=U(s)\bigotimes_{j=1}^{n}P_{j}(0)U^{\dagger}(s),
\end{equation}
where
\begin{equation}\label{Us}
U(s)=\bigotimes_{j=1}^{n}A_{j}(s).
\end{equation}
We suppose that the eigenvalue of the above projection is $hLK(s)$
and $m$ is an additional quantum number to distinguish the
degenerate eigenspaces belonging to the eigenvalue. The Zeno
subspace and the corresponding projection belonging to the the
eigenvalue $hLK(s)$ are respectively
\begin{eqnarray}
{\cal H}_{P_{L}}(s)&=&\bigoplus_{m}{\cal H}_{P_{Lm}}(s),\\
P(L, s)&=&\sum_{m}P(L m,s).
\end{eqnarray}
We assume that the magnitude $h$ of the field is larger than the
critical point $h_{c}=4$ \cite{Korepin16} and the initial state of
the measured system is the ground state(ferromagnetic):
\begin{equation}
|n,0\rangle=\bigotimes_{j=1}^{n}\pmatrix{ 1 \cr 0\cr}_{j}.
\end{equation}
Since the ground state is non-degenerate, we denote the
corresponding projection by P(n,0). Using Eq.(\ref{JP}), the jump
probability from Zeno subspace ${\cal H}_{P_{n}}(0)$ to ${\cal
H}_{P_{L}}(s)$ is
\begin{eqnarray}\label{Wn}
W(P(n,0)\rightarrow P(L
,1))&=&\int_{0}^{1}ds_{1}\int_{0}^{1}ds_{2}Tr\{U^{\dagger}(s_{1})H_{0}(s_{1})U(s_{1})P(n,0)
U^{\dagger}(s_{2})H_{0}(s_{2})U(s_{2})P(L,0)\} \nonumber \\
                          &&\times\exp\{i\int_{s_{2}}^{s_{1}}h T(n-L)K(s^{'})ds^{'}\}.
\end{eqnarray}

Now for simplicity, let us consider two qubits described by free
Hamiltonian $H_{0}$
\begin{equation}\label{H0}
 H_{0}=\sigma_{1}^{x}\sigma_{2}^{x}+2\sigma_{1}^{y}\sigma_{2}^{y}+\sigma_{1}^{z}\sigma_{2}^{z}.
 \end{equation}
$H_{0}$ can be regard as a perturbation in comparison with $I(s)$
as long as we tune the magnitude $h$ of the field. We denote the
projections onto the eigenspaces  of $I(s)$ by
\begin{eqnarray}
P_{1}(s)&=&P(\uparrow\uparrow,s) \quad\quad
P_{2}(s)=P(\uparrow\downarrow,s) \nonumber \\
P_{3}(s)&=&P(\downarrow\uparrow,s) \quad \quad
P_{4}(s)=P(\downarrow\downarrow,s),
\end{eqnarray}
which belong to the eigenvalues $(-2hK(s))$, 0, 0, $(2hK(s))$,
respectively. Therefore the Zeno subspaces belonging to the
eigenvalues $(-2hK(s))$, 0, $(2hK(s))$ are respectively
\begin{eqnarray}
{\cal H}_{P_{1}}(s),\quad {\cal H}_{P_{2}}(s) \bigoplus {\cal
H}_{P_{3}}(s),\quad {\cal H}_{P_{4}}(s).
\end{eqnarray}
At $t$=0, we have
\begin{eqnarray}
  P_{1}(0)&=&\pmatrix{ 1 & 0 \cr 0 & 0\cr }_{1} \pmatrix{1 & 0 \cr 0 & 0 \cr }_{2} \quad
  \quad \quad
  P_{2}(0)=\pmatrix{ 1 & 0 \cr 0 & 0\cr }_{1} \pmatrix{0 & 0 \cr 0 & 1 \cr }_{2} \nonumber \\
  P_{3}(0)&=&\pmatrix{ 0 & 0 \cr 0 & 1 \cr }_{1} \pmatrix{1 & 0 \cr 0 & 0\cr }_{2}
  \quad\quad \quad
  P_{4}(0)= \pmatrix{ 0 & 0 \cr 0 & 1\cr }_{1} \pmatrix{0 & 0 \cr 0 & 1\cr
  }_{2}.
  \label{P0}
  \end{eqnarray}
From Eq.(\ref{A}), (\ref{AD}), (\ref{Us}), (\ref{Wn}), (\ref{H0})
and (\ref{P0}), We find the jump probability has the simple form
\begin{eqnarray}
 W(P_{1}(0)\rightarrow P_{2}(1))&=&0,   \\
 W(P_{1}(0)\rightarrow P_{3}(1))&=&0,  \\
 W(P_{1}(0)\rightarrow P_{4}(1)) \nonumber
      &=&\int_{0}^{1}ds_{1}\int_{0}^{1}ds_{2}\exp(i\int_{s_{2}}^{s_{1}}4hTK(s^{'})ds^{'})\nonumber\\
      &=&\int_{0}^{1}ds_{1}\int_{0}^{1}ds_{2}\cos(4hT\int_{0}^{s_{1}}K(s^{'})ds^{'}-4hT\int_{0}^{s_{2}}K(s^{'})ds^{'})
      \nonumber \\
      &=&(\int_{0}^{1}ds\cos(4hT\int_{0}^{s}K(s^{'})ds^{'}))^{2}+(\int_{0}^{1}ds\sin(4hT\int_{0}^{s}K(s^{'})ds^{'}))^{2}.
      \label{final}
 \end{eqnarray}
We can see that the jump probability from ${\cal H}_{P_{1}}(s)$ to
${\cal H}_{P_{2}}(s) \bigoplus {\cal H}_{P_{3}}(s)$ is zero, which
can be explained by the following matrix element:
\begin{eqnarray}
\langle\uparrow\downarrow,
s|H_{0}|\uparrow\uparrow, s\rangle&=&0. \label{92} \\
\langle\downarrow\uparrow, s|H_{0}|\uparrow\uparrow,s\rangle&=&0.
\label{93}\\
\langle\downarrow\downarrow,
s|H_{0}|\uparrow\uparrow,s\rangle&=&-1. \label{94}
\end{eqnarray}
From Eq.(\ref{92}) and (\ref{93}), we see that jump process from
${\cal H}_{P_{1}}(s)$ to ${\cal H}_{P_{2}}(s) \bigoplus {\cal
H}_{P_{3}}(s)$ is forbidden under the action of $H_{0}$ at any
time. Therefore the jump probability running out of ${\cal
H}_{P_{1}}(s)$ is the contribution of the probability
$W(P_{1}(0)\rightarrow P_{4}(1))$. Comparing the matrix elements
of $H_{0}$ with the minimal energy difference $2hK(s)$ of $I(s)$,
we also find that $H_{0}$ is a perturbation with the condition
$h\geq4$. On the other hand, from the result of
Eq.(\ref{adiabatic}), we get the adiabatic approximation condition
$hT\gg1$. The above two conditions are for the validity of
Eq.(\ref{final}).

Fig.1 shows the jump probability  which varies with the time T in
the form $sin^{2}T$ without the interaction $I(s)$. Fig.2 and
Fig.3 show two cases of the jump probability determined by
Eq.(\ref{final}). It is given in Fig.2 that the probability of the
system jumping out of ${\cal H}_{P_{1}}(s)$ varies with the
magnitude $h$ of the field, where the duration T of the
measurement is 1 and the magnitude is larger than 9 for the
adiabatic approximation condition. We see that the amplitude of
the probability declines with the enhancement of the magnitude. In
Fig.3, we change the duration of the measurement from 1 to 10 with
$h=9$. The amplitude tends to zero rapidly with the duration $T$
increasing. Therefore we find that the measurement does slow down
the decay of ${\cal H}_{P_{1}}(s)$ by enhancing the magnitude of
the field or the duration of the measurement. The quantum Zeno
effect takes place.

\begin{figure}
\begin{center}\leavevmode
\epsfxsize=8cm\epsfbox{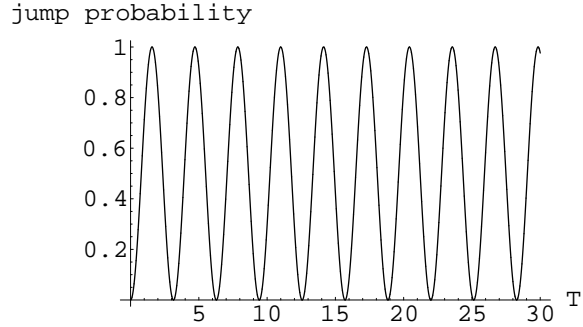}
\end{center}
\caption{the jump probability running out of ${\cal H}_{P_{1}}(s)$
varies with the time T without the action of the field. The
amplitude does not change with T increasing.}
\end{figure}
\begin{figure}
\begin{center}\leavevmode
\epsfxsize=8cm\epsfbox{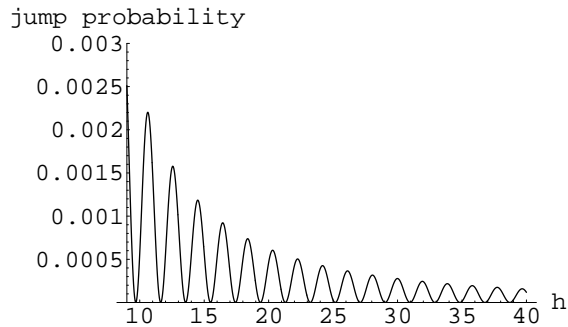}
\end{center}
\caption{The jump probability running out of ${\cal H}_{P_{1}}(s)$
varies with the magnitude h of the magnetic field, where the
duration of the measurement $T=1$. The amplitude declines with h
increasing.}
\end{figure}
\begin{figure}
\begin{center}\leavevmode
\epsfxsize=8cm\epsfbox{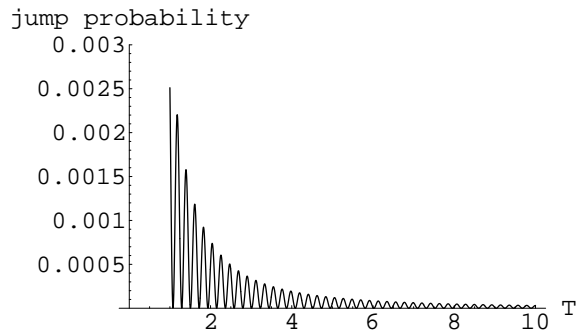}
\end{center}
\caption{The jump probability running out of ${\cal H}_{P_{1}}(s)$
 varies with the duration of the measurement,
 where the magnitude of the magnetic field $h=9$.
 The more slowly we rotate the magnetic field the smaller
 the amplitude is.}
\end{figure}

\section{Conclusions}

\indent

We has analyze the time-dependent measurement and get a general
expression(\ref{JP}) of jump probability between different Zeno
subspaces of the interaction Hamiltonian $KH_{meas}(s)$. The
validity of the expression has two conditions: the free
Hamiltonian can be regarded as a perturbation of $KH_{meas}(s)$
and  $KH_{meas}(s)$ changes efficiently slowly to satisfy the
adiabatic approximation condition. Therefore the result is a
general perturbative method which describe the dynamics of quantum
Zeno subspaces. It can be applied to not only the time-independent
measurement, but also the time-dependent one. We use this
expression in two time-independent measurement's examples to
explain the quantum Zeno effect. We also use it in a
time-dependent measurement on an XYZ Heisenberg spin chain. In
this measurement the Zeno subspace we adopt is one-dimensional. We
can see that the jump probability out of its initial Zeno subspace
reduces as the coupling constant $K$. In fact, the analysis of the
multidimensional Zeno subspace is just similar with that of the
one-dimensional's. Now we are able to apprehend the dynamics of
the quantum Zeno subspaces more.

It is well known that it is important that one can prepare and/or
control the state of the system under consideration at one's will
in quantum information and computation. Recently, a novel
mechanism to purify quantum states, based on the Zeno-like
measurements, has been proposed \cite{Nakazato}. The purification
process of states characterized by the specific interactions of
the systems was shown \cite{Militello} to be controlled through
the continuous measurements, i.e., the quantum Zeno dynamics. We
believe that the perturbative approach for the quantum Zeno
dynamics given by us here is helpful to discuss the quantum state
purification.
\\
\\
\\
\textbf{Acknowlegments}
\\
\\

The work was partly supported by the NNSF of China (Grant
No.90203003), NSF of Zhejiang Province (Grant No.602018), and by
the Foundation of Education Ministry of China (Grant
No.010335025).

\end{document}